\documentclass[iop]{emulateapj}

\begin{document}

\title{Radio flaring from the T6 dwarf WISEPC J112254.73+255021.5 with a possible ultra-short periodicity}

\author{Matthew Route\altaffilmark{1,2,3} \& Alexander Wolszczan\altaffilmark{1,2}}

\altaffiltext{1}{Department of Astronomy and Astrophysics, the Pennsylvania State University, 525 Davey Laboratory, University Park, PA 16802, alex@astro.psu.edu}

\altaffiltext{2}{Center for Exoplanets and Habitable Worlds, the Pennsylvania State University, 525 Davey Laboratory, University Park, PA 16802}

\altaffiltext{3}{Current Address: Research Computing, Information Technology at Purdue, Purdue University, 155 S. Grant St., West Lafayette, IN 47907, mroute@purdue.edu}

\slugcomment{Accepted for publication in ApJL; 2016 March 27}

\begin{abstract}
We present new results from a continuing 5 GHz search for flaring radio emission from a sample of L and T brown dwarfs, conducted with the 305-m Arecibo radio telescope. In addition to the previously reported flaring from the T6.5-dwarf 2MASS J10475385+212423, we have detected and confirmed circularly polarized flares from another T6-dwarf, WISEPC J112254.73+255021.5. Although the flares are sporadic, they appear to occur at a stable period of 0.288 hours. Given the current constraints, periods equal to its second and third subharmonic cannot be ruled out. The stability of this period over the 8-month timespan of observations indicates that, if real, it likely reflects the star's rapid rotation.  If confirmed, any of the three inferred periodicities would be much shorter than the shortest, 1.41-hour rotation period of a brown dwarf measured so far. This finding would place a new observational constraint on the angular momentum evolution and rotational stability of substellar objects. The detection of radio emission from the sixth $\sim$1000 K dwarf further demonstrates that the coolest brown dwarfs and, possibly, young giant planets, can be efficiently investigated using radio observations at centimeter wavelengths as a tool.  \end{abstract}
 
\section{Introduction}
Radio observations of ultracool dwarfs (UCDs; \citep{kir97}) address important issues in the physics of low-mass stars, brown dwarfs,  and planets. The most commonly discussed problem is the generation and topology of the magnetic fields in fully-convective, low-mass stars and brown dwarfs \citep{don06,mor10}. Because these objects are cool, rapid rotators, the usual magnetic activity indicators provided by H$\alpha$ and X-ray emission become inefficient, and the Zeeman splitting based technique does not work, due to the rotational broadening of spectral lines employed in this method \citep{giz00,ste06,rei07,mcl12}. On the other hand, radio detections, especially of the electron cyclotron maser (ECM) driven, rapid bursts of emission, enable a direct measurement of the lower bound to the local magnetic field strength (\citet{tre06}, and references therein), provide means to measure stellar rotation \citep{wilb15}, and diagnose the conditions of the magnetosphere - ionosphere interaction \citep{nic12}. In addition, modeling of the time-frequency (dynamic) spectra of these bursts provides information on the UCD magnetic field topology \citep{hes11,kuz12,lyn15}.

Another problem that can be addressed using radio detection as a tool is the angular momentum evolution of UCDs \citep{zap06,rei08}. Rotation period measurements of young brown dwarfs in clusters demonstrate that these objects lose very little of their initial angular momentum as they evolve \citep{sch15}. Measurements of the periodic radio bursts generated by the coolest, old dwarfs probe the late stages of their rotational evolution. The advantage of using ECM-generated radio flaring for period measurements over optical/near-infrared photometry is that their precision is not dependent on cloud-induced intensity variations. For example, the apparent radio flaring period of a rapidly-rotating, M9-type UCD, TVLM 513-46546, shows only a $\sim$0.03\% variation around the average value of 7054.468 s calculated over a 7-year baseline \citep{wol14}. In contrast, the measurability and precision of photometrically-derived periods critically depend on the ratio of the period itself to the timescale of cloud evolution \citep{hei15}.

The late L, T, and Y-type dwarfs detectable in the radio can also be treated as proxies for studies of the magnetic properties of young, massive exoplanets, assuming, for example, the magnetic field scaling predictions discussed by \citet{rc10}. Although attempts to detect low-frequency radio emission from hot Jupiters generated by the stellar wind - planetary magnetosphere interaction have not been successful (e.g. \citet{laz10}, and references therein), GHz-frequency measurements of hot exoplanets discovered by direct imaging at optical wavelengths (e.g. \citep{mar08}) may offer a practical alternative in the form of detections of ECM radio bursts similar to those generated by the coolest UCDs \citep{rou12,kao16}.

Motivated by these considerations, we have been searching for polarized radio flaring from these objects with the Arecibo radio telescope. Our first survey of 33 UCDs at 5 GHz has resulted in the first discovery of a radio emitting T-dwarf, the T6.5 UCD, 2MASS J10475385+212423 (J1047+21; Route \& Wolszczan 2012, 2013). Follow-up Jansky Very Large Array (JVLA) observations detected quiescent radio emission from this object \citep{wil13}, confirmed its flaring component \citep{wilb15,kao16}, and produced a measurement of the 1.77-hr rotation period of the star \citep{wilb15}. In addition, guided by the fact that J1047+21 is also a weak H$\alpha$ emitter, \citet{kao16} have detected quiescent and flaring radio emission from four out of five H$\alpha$ and optical/infrared variability-selected L and T-type UCDs.

Here, we report the Arecibo detection of another T-dwarf, WISEPC J112254.73+255021.5 (J1122+25), in the course of the second part of our unbiased survey of late L and T-type UCDs. The most distinguishing characteristic of this new radio-emitting UCD is its tentatively established ultra-fast rotation period. If confirmed, it will become by far the shortest one measured for any UCD. 

Our observations, the detection of flares from J1122+25, their characteristics, and a tentative measurement of the UCD's rotation period are described in Section 2. Further discussion of these results and our conclusions are given in Section 3. The entire Arecibo UCD survey will be described in a forthcoming paper.

\section{Detection of Flaring from WISEPC J112254.73+255021.5}

WISEPC J112254.73+255021.5 was discovered in the course of the Wide-field Infrared Survey Explorer survey (WISE; Kirkpatrick et al. 2011).  Medium resolution near-infrared spectroscopy using the SpeX on NASA Infrared Telescope Facility (IRTF) led to the classification of the object as a T6 brown dwarf, with a spectrophotometric distance estimate of 16.9 pc.  Nearby M5 field star LHS 302 has similar distance and proper motion to that of J1122+25, but due to its low mass and a wide projected separation of $\sim$4500 AU, the objects are unlikely to be physically bound.  

\footnotetext[1]{ ``The VLA FIRST Survey,'' available at http://sundog.stsci.edu/index.html.}

The continuing L and T- spectral type dwarf survey with the Arecibo radio telescope uses the instrumental setup and data analysis method discussed in detail by \citet{rou12,rou13}. J1122+25 has been detected on 2013 May 8 as a broadband, left-hand circularly polarized (LCP), $\sim$1.5 mJy flare. Four additional LCP flares have been identified in the data taken on 2013 December 27 and 31, and 2014 January 4, during a weeklong confirmation run. The 2013 December 27 event, shown in Fig. 1 as the most prominent example of the J1122+25 flaring, consists of two 100\% LCP bursts, occurring $\sim$1100 s apart, with peak fluxes of $\sim$3 mJy and $\sim$2.5 mJy, respectively, at the full, 0.1 s resolution. An additional 10 hours of observations made in 2015 May and June, non-simultaneously covering the 3-4 GHz and 4-5 GHz bands, have not produced any additional detections. A search of the ``14Dec17'' release of the FIRST radio catalog\footnotemark ~found no radio sources within the half-power beam width of the telescope, with a detection limit of 0.92 mJy. The nearest sources are the star J112253.4+255250.2 at a separation of 148.8 arcsec, and J112247.6+254736.9, an unknown object 184.3 arcsec away with peak fluxes of 1.19 mJy and 4.08 mJy, respectively, at 1.4 GHz. These sources are located well outside of the Arecibo $\sim$1 arcmin beam at 5 GHz.

\subsection{Flare Characteristics}

J1122+25 is only the second radio emitting T-dwarf identified in an unbiased survey, after our detection of flaring from J1047+21 \citep{rou12}. J1122+25 emits broadband, short duration, 30 - 120 s pulses, characterized by a rapid, variable frequency drift (50-800 MHz s$^{-1}$) and left circular polarization ranging from 15\% to 100 \%. These characteristics are typical of ECM emission generated by electrons moving in the magnetic fields of rapidly rotating UCDs \citep{bur05,hal06,hal08,ber09,rou12,wilb15,wil15}. Similar to J1047+21, the emission is clearly sporadic with only five pulse detections in 29 hours of observations. However, as observed by \citet{wilb15}, J1047+21 does undergo periods of activity, during which several varying, consecutive pulses can be detected.

The shortest measurable timescale of intensity fluctuations in the flares can be used to estimate an upper limit to the source size and a lower limit to its brightness temperature. We have computed a noise-corrected autocorrelation function (ACF) of the strongest, December 27 flare sampled at the original, 0.1 s resolution, and integrated over the interference free parts of the total bandpass, $\Delta\nu$=1 GHz (Fig. 2). The ACF shows a clear change of slope around the $\sim$2 s lag, indicating the presence of intensity fluctuations in the flare on a similar timescale. As the flare rapidly drifts in frequency at $d\nu/dt\sim$50--800 MHz s$^{-1}$, this timescale most likely represents intensity variations that are faster than our 0.1 s resolution, but they are smeared by frequency drifting to $\Delta\nu(d\nu/dt)^{-1}\sim$1.25--2 s. This places an upper limit of $\sim$0.4 R$_J$ on the size of the emitting region, and leads to the source's brightness temperature estimate, T$_b\geq 4\times 10^{11}$ K.

The high degree of circular polarization of the J1122+25 flares and the above T$_b$, which is close to the T$_b\sim10^{12}$ K Compton limit of compact radio sources \citep{kel69}, make it entirely feasible to assume that the ECM mechanism is responsible for their generation. In this case, the star's local magnetic field strength, $B$, can be estimated from the corresponding cyclotron frequency, $\nu_{c} = 2.8 \times 10^{6}$ B (Gauss), where $\nu_{c} = 5.2 \times 10^{9}$ Hz, represents the upper end of the 1 GHz receiver bandpass. Because the observed emission appears to extend beyond that frequency for all five flares, the cutoff frequency of emission has yet to be measured, and the computed $B\gtrsim1.8$ kG must be treated as a lower limit to the actual magnetic field strength of the star. 

\footnotetext[2]{Information available at http://tempo.sourceforge.net/.}

\subsection{Periodicity in the J1122+25 Flaring?}

We have investigated the possibility that, despite their sporadic character, the flares may be periodic. If all five flares appear in phase with the dwarf's rotation period, it is obviously constrained by the $\sim$1100 s separation of the two December 27 events (Fig. 1). To refine this period estimate and test whether the five flares are indeed phase-connected, we have used a Monte Carlo (MC) approach in which trial periods have been randomly drawn from a 15 to 20 min. interval, and fitted to flare times-of-arrival (TOAs) with the TEMPO\footnotemark ~modeling code in a manner described in detail by \citet{wol14}.

The modeling process is complicated by the high variability of the flare profiles, thus making it impossible to unambiguously determine the phase reference points for TOA measurements. Therefore, in our procedure, we have included a random selection of TOAs from windows which were centered on the flares and spanned the full widths of flare profiles at the 10\% flux density level. TEMPO was used to fit 2000 trial periods to each of the 500 sets of randomly-selected, five flare TOAs. Surprisingly, we have easily obtained a phase-connected timing solution with a new, improved period of 1035.7881$\pm$0.0008 s and a post-fit rms residual, $\sigma\approx$0.86 s. Phase alignment of the five flares at the best-fit period is shown in Fig. 3. The same period is obtained with the flare TOAs fixed at the approximate centers of the profiles, but the post-fit $\sigma$ becomes $\sim$10 s, which reflects the TOA uncertainty due to the changing flare morphology (see also \citep{wol14}).

As the result of period modeling based on only five data points must be treated with caution, we have checked its statistical significance with a MC simulation in which we have randomly drawn 5$\times$10$^5$ sets of five time-tags of samples from the five, 10-minute, 667-data point observing windows containing flares (Figs 1,3), and used TEMPO to fit the measured best-fit period to these simulated arrival times. By counting the post-fit $\sigma\leq$10 s values from these fits, we have computed a 0.01\% false alarm probability (FAP) for our period measurement of the J1122+25 flares. For a much more restrictive, $\sigma\leq$1 s criterion, the FAP becomes vanishingly small.

Another possibility is that the two, apparently consecutive, December 27 flares were generated by different activity regions in the magnetosphere, and only one of them was rotational phase-connected to the remaining three events. Such behavior has been observed in other well-studied, periodically radio emitting UCDs, such as TVLM 513 and 2MASS J07464256+2000321 (\citet{lyn15}, and references therein). We have therefore conducted MC tests in the manner described above by removing either of these two flares from the data set. The best-fit period for the four flares with the first December 27 one removed is 1032.6988$\pm$0.0008 s with $\sigma$=0.16 s. Not surprisingly, this is very close to the 1035 s period calculated for the five flare case. The data are also very well fitted with the approximately third subharmonic of that period, 3107.519$\pm$0.002 s and $\sigma$=0.82 s. When the second flare is excluded, the best-fit period is 2053.1398$\pm$0.0005, $\sigma$=0.22 s, expectably close to twice the shortest period, which necessarily produces the same fit. The FAP values for the two fits, using the $\sigma\leq$10 s criterion, are 0.2\% and 0.02\%, respectively.  As discussed above, it is important to remember that the tiny, formal TEMPO errors associated with the best-fit periods are much smaller than the actual, more realistic uncertainties of $\sim\pm$15 s resulting from the flare profile variability and the associated ambiguity in determining the actual TOAs. These uncertainties have been estimated from the scatter of the best-fit periods with $\sigma\leq$10 s around the one with the lowest $\sigma$ value.

Finally, we have folded all the available data not containing the five flares, modulo the best-fit periods, to check for the possible presence of low-level, periodic emission. This analysis has not produced any obvious detection, down to the 3$\sigma$ level of $\sim$30 $\mu$Jy. We conclude that both the available observational evidence and the FAP estimates suggest that one of the three, obviously harmonically-related periodicities in the J1122+25 flaring over the period of 8 months is very likely to be real, but additional, extensive, possibly multi-wavelength observations are needed to verify this result.

\section{Discussion and Conclusions}

Our detection of radio flaring from the T-dwarf J1122+25, reported in this paper, significantly adds to the small group of five late-L and T-dwarfs previously known to display this type of activity \citep{rou12,kao16,wil15}. The J1122+25 flares are LCP with brightness temperatures approaching 10$^{12}$ K, indicating their ECM origin. In this case, the local magnetic field strength must exceed 1.8 kG, in agreement with similar measurements of other radio detected UCDs \citep{bur05,hal06,hal08,ber09,rou12,wil15}. 

Clearly, the most interesting aspect of our radio detection of J1122+25 is the tantalizing possibility that the flares from this UCD occur at a period that may be as short as $\sim$17 minutes. The shortest rotation periods derived from the photometric variability of UCDs at optical/near-infrared wavelengths are 1.41 and 1.55 hours, for the T6.5 dwarf 2MASS J22282889-4310262 \citep{cla08} and the T7 dwarf 2MASS J00501994-3322402 \citep{met15}, respectively. An even shorter, but never confirmed periodicity of 16.5 min has been reported for the T4.5 dwarf 2M 0559-1404 \citep{koen04}. \citet{bai01} and \citet{zap03} have observed a varying, 30-46.4 min periodicity in the young, M8.5 dwarf S Ori 45, but, obviously, even if real, it could not be interpreted in terms of stellar rotation, because of its variability.

In principle, the rotational breakup period of an old brown dwarf, $P_0 = 2\pi(R^3/(GM))^{1/2}$, could be as short as $\sim$20 min, using the standard value of R, which is the Jupiter radius, R$_J$=7$\times$10$^7$ m, and the mass, M$\leq$80 M$_J$. Assuming $P_0=1035$ s, the shortest of the three possible periods of the J1122+25 flares, one needs R$\leq$0.9 R$_J$ to ensure that $M\leq80 M_J$. Taking into account a significant oblateness of the star induced by rapid rotation, and using the Darwin-Radau relationship \citep{bar03,sen10}, the oblateness of a rotating star is given in terms of its polar and equatorial radii, R$_p$ and R$_e$, as:
$$f=1-{R_p\over R_e}={{\Omega^2R\over g} {\left[{5\over 2}\left(1-{3K\over 2}\right)^2 + {2\over 5}\right]}^{-1}},$$
where $\Omega$ is the angular rotational velocity of the star, K=I/(MR$^2$) ranges from 0.261 to 0.205 for the polytropic indices n=1 and n=1.5, respectively, and I is the moment of inertia. Assuming the observed value of $\Omega$=6.1$\times$10$^{-3}$ s$^{-1}$, n=1.5 for high surface gravity, the corresponding breakup oblateness, $f=0.38$ \citep{jam64},  and R=0.8 R$_J$, we get log(g)$\sim$5.5. This, together with constraints provided by the dwarf's temperature of $T_e\sim1000$ K \citep{kir11}, and the evolutionary models of \citet{bur01}, gives M$\geq$60 M$_J$ and a lower limit of $\sim$1 Gyr to the age of the star.

Of course, the two longer periods produce correspondingly less extreme characteristics of the dwarf. For example, for P$_0$=2053 s and R=R$_J$, the Darwin-Radau approximation gives log(g)$\geq$5, which leads to M$\geq$35 M$_J$ for T$_e$=1000 K. If any of the three measured periods is real, all these estimates make J1122+25 an ultra-rapidly rotating, massive UCD with the radius falling well within the theoretical range for dwarfs older than 1 Gyr. 

Despite the plausibility of the existence of sufficiently evolved dwarfs rotating at periods $\leq$1 hour, based on the investigations of the angular momentum loss in the evolving substellar objects (see \citet{bou14}, for a review), none have been detected so far.  Whereas objects rotating close to their breakup periods may be rare, this situation could also be caused by observational biases. However, several ground and space-based surveys of L and T dwarfs conducted in the past had sufficiently high sensitivity and cadence of photometric measurements to be able to detect such periodicities (e.g. \citet{met15}, and references therein). On the other hand,  photometric fluctuations of the UCDs generated by cloud dynamics can dramatically blur their rotation-induced variability, even for the shortest periods observed so far \citep{rad12,apai13,hei15}. The same could happen for objects with rotational periods $\leq$1 hour, if the ``weather'' related variations on such the short timescales are still feasible.

Ideally, assuming any of the three possible periods discussed above, and a quasi-steady emission source, flares from J1122+25 should be detectable at least once during each, typically $\sim$2-hour long observing session. Flaring from this dwarf is clearly sporadic, but, as in the case of the first radio detected T-dwarf, J1047+21 \citep{rou12}, it could occasionally undergo periods of higher activity \citep{wilb15}. The sporadic appearance of flares from J1122+25 and a few other radio-emitting T-dwarfs detected so far is most likely related to variations in the flow of electrons into the magnetic field needed to power the observed, ECM-generated flaring.  Given the star's largely neutral, field-decoupled atmosphere, \citet{moh02}, \citet{sch09}, and \citet{nic12} have argued that, like in the cases of Jupiter and Saturn, the power needed to create such currents in the magnetosphere is extracted from rotation. In particular, \citet{nic12} propose that such currents could be created through magnetic field reconnection at the boundary between the closed and the open field lines, where the magnetosphere opens up to the interstellar medium. As the conditions in this transition region are expected to be highly variable, it would naturally explain the intermittent character of the observed emission.

Of course, more work is needed to confirm the J1122+25 flaring period reported in this paper. This includes a study of its photometric variability and, if possible, a measurement of the $v~sin~i$ value, which, if the rotation period reported here is real, is likely to be much above 100 km s$^{-1}$. More attempts to detect radio flaring from the dwarf are also important. Finally, it would be of interest to verify if J1122+25 is a detectable H$\alpha$ emitter and/or it exhibits any enhanced variability at optical/infrared wavelengths. If so, it would further strengthen the evidence presented by \citet{kao16} that the best candidates for detectable radio emission among the coolest UCDs are the ones that exhibit one or both of these properties.
 
\section{Acknowledgements}

MR acknowledges support from the Center for Exoplanets and Habitable Worlds, which is supported by the Pennsylvania State University, and the Eberly College of Science. The Arecibo Observatory is operated by SRI International under a cooperative agreement with the National Science Foundation (AST-1100968), and in alliance with Ana G. M\'{e}ndez-Universidad Metropolitana, and the Universities Space Research Association. This paper has made use of NASA's Astrophysics Data System.

\begin{figure}
\centering
\includegraphics[width=1.05\textwidth,angle=0]{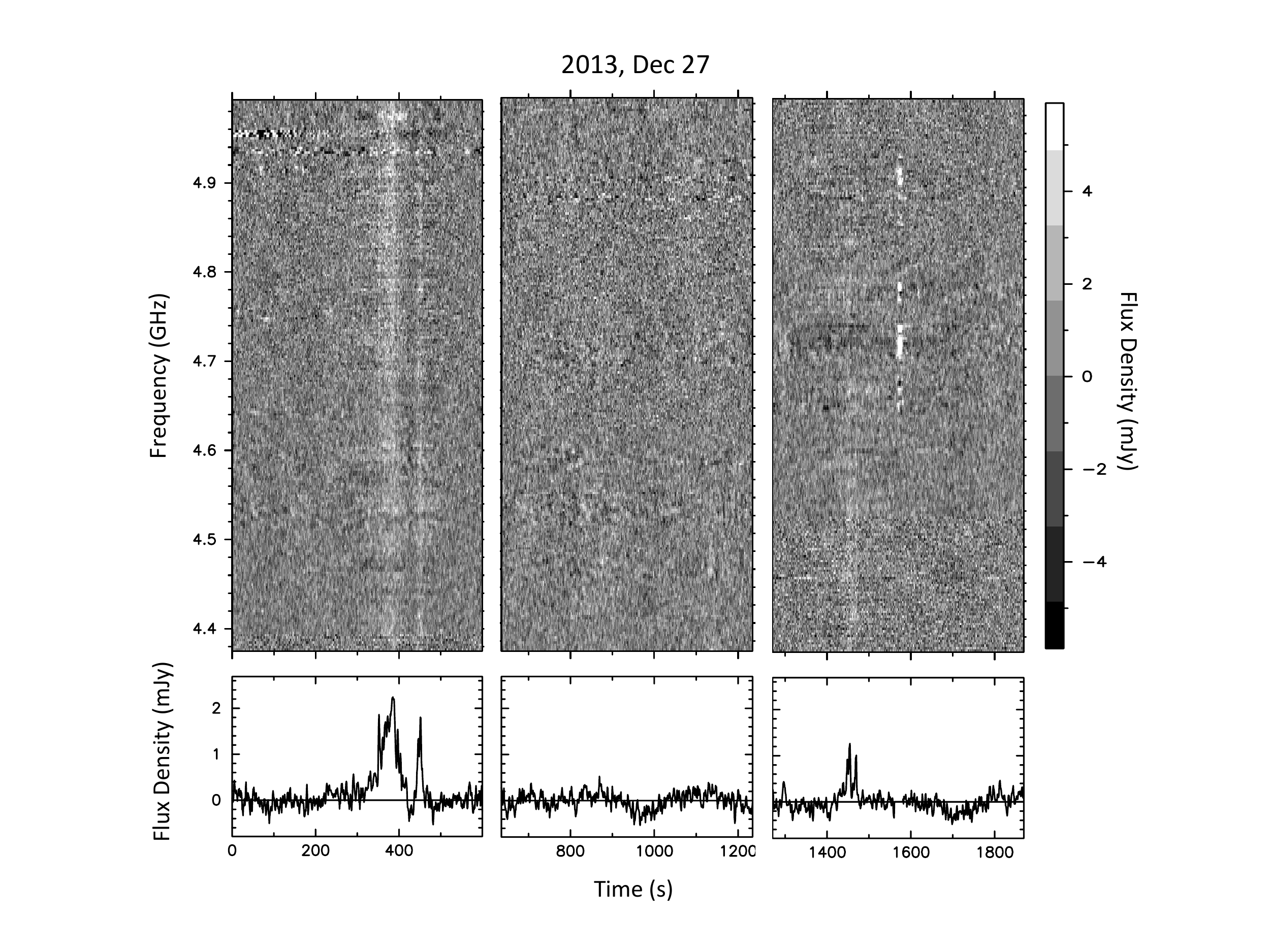}
\caption{Dynamic spectra (top) and average time profiles (bottom) of the two LCP flares from J1122+25 displayed at the time and frequency resolution of 2.7 s and 2.5 MHz, respectively. The 35 s gaps between the adjacent scans are times spent on calibration. A series of spikes at $\sim$1600 s, centered at $\sim$4.75 GHz, is due to interference of a terrestrial origin.  \label{fig1}}
\end{figure}

\begin{figure}
\centering
\includegraphics[width=0.8\textwidth,angle=-90]{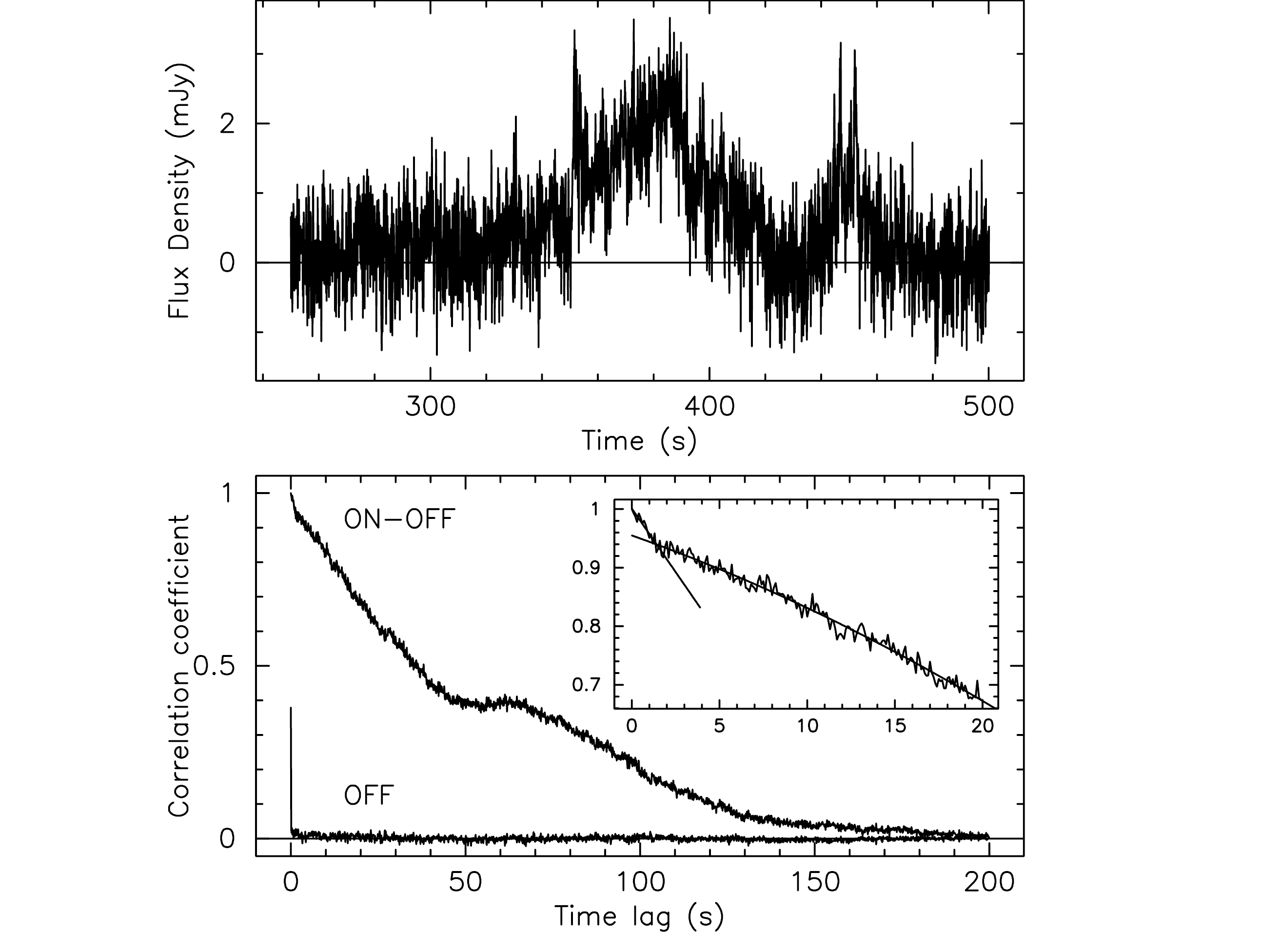}
\caption{(Top) Bandpass-averaged profile of a circularly polarized flare from J1122+25 observed on 2013 December 27 displayed at the full, 0.1 s time resolution. (Bottom) A noise-corrected ACF of the flare profile. The inset shows the first 200 lags of the ACF with the solid curves denoting polynomial fits made to emphasize the ACF slope change at the lag of $\sim$2 s. \label{fig2}}
\end{figure}

\begin{figure}
\centering
\includegraphics[width=1.1\textwidth,angle=0]{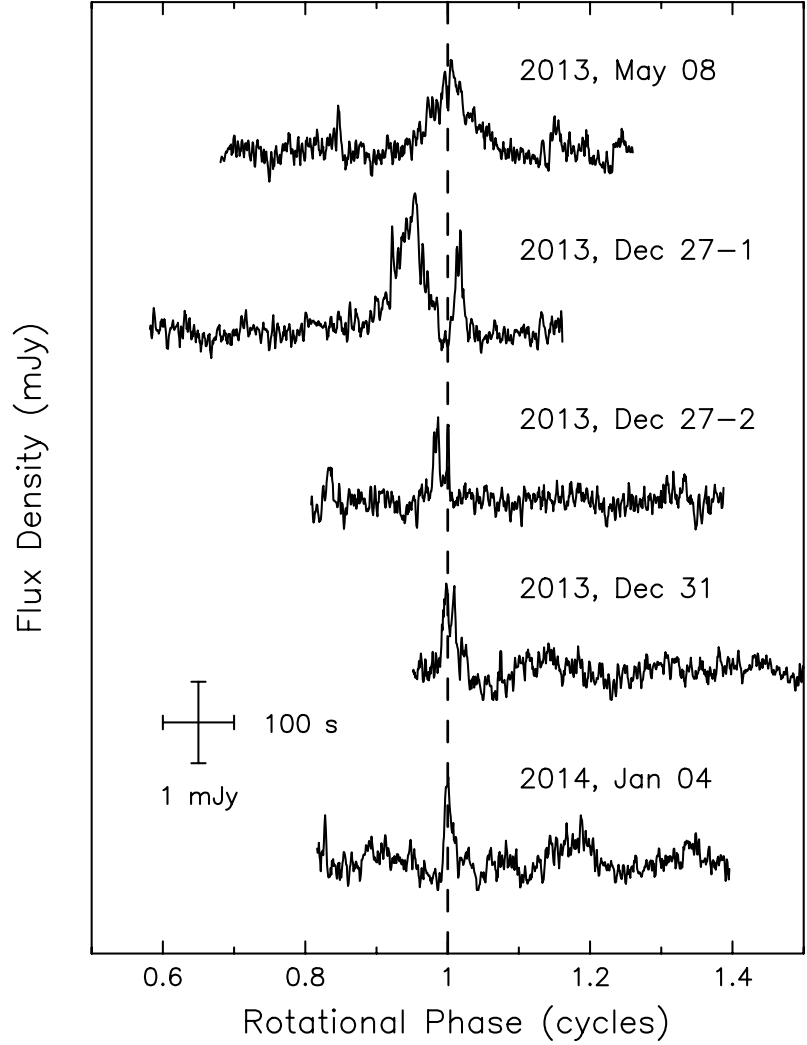}
\caption{The arrival times of the five flares from J1122+25 observed between 2013 May and 2014 January. The flares are bandpass-integrated, smoothed to the 2.7 s resolution, and phase-aligned assuming the best-fit, 1035 s period. The vertical, dashed line marks the arrival times predicted by the model. The actual, best-fit flare arrival times selected by the MC simulation process are: MJD 56421.05858, 56653.39151, 56653.40348, 56657.35930, and 56661.35106.\label{fig3}}
\end{figure}

\end{document}